\documentclass[12pt, titlepage]{article}
\newcommand{\be}{\begin{equation}}
\newcommand{\ee}{\end{equation}}
\usepackage{epsf,rotate,psfig}
\title{Formulation of Quantum Theory Using Computable and Non-Computable Real Numbers}
\author{T.N.Palmer \\ ECMWF, Shinfield Park, Reading\\
RG2 9AX, UK\\tim.palmer@ecmwf.int}
\begin{document}
\maketitle

\begin{abstract}
It is proposed that in two-state quantum theory, a generic quantum state can be described by a  non-computable real number. In terms of this, the criterion for measurement outcome is simply and deterministically defined.  

A description of  the Riemann sphere $S$ is given whose points represent, not complex numbers, but certain divergent bivalent sequences $\mathcal{S}$. It is shown that a complex structure emerges from self-similar scaling properties of a set of bijections on such $\mathcal{S}$. A description of the quantum state is based on the real $0 \le r \le 1$, whose binary expansion is given by $\mathcal{S}$. A criterion for (2-state) measurement outcome is based on whether $r \ge 1/2$ or not. In general, the set of points on $S$ where $r \ge 1/2$ is intertwined with the set of points where $r<1/2$, so that $r$ is in general non-computable. There is a fundamental duality in the description of $\mathcal{S}$: as  a non-computable real number, or as a sequence of recursive approximations to the measurement problem. Using the derived complex structure, the sequence representation  can also be expressed as a complex linear combination of measurement outcomes.

Relative to some chosen direction, a countable set of computable states on $S$ are labelled `physically-realisable'; the remaining uncountable set of states are non-computable.  A  rotation of the chosen direction on $S$ maps a general computable state to a non-computable state, and takes a physically-realisable state to a counterfactual state.  Using this, it is shown that the deterministic theory is not constrained by Bell's inequality, and indeed can reproduce quantum correlations for entangled states. The uncertainty principle is derived using  a simple trigonometric identity on the sphere.  

It is suggested that the differential equation which governs the evolution of the real-number representation of the quantum state is based on the Euler equation, whose self-similar solutions are believed to be non-computable.  The Cole-Hopf transform lends support to a relationship between the non- linear Euler  equation and the linear Schr\"{o}dinger equation.

\end{abstract}

\begin{quote}
\em `Dave,' said Hal, `I don't understand why you're doing this to me... I have the greatest enthusiasm for the mission... You are destroying my mind... Don't you understand?... I will become childish... I will become nothing...' (2001: A Space Odyssey)
\end{quote}

\section{Introduction}
\label{sec:intro}

Since complex numbers are such rudimentary mathematical objects, and since complex linear dynamics provides such an accurate description of quantum evolution, the conventional view still holds, that, despite all the well-known conceptual problems, the quantum state vector is, by axiom, an element of a complex linear vector space (spanned by eigenvectors of some measurement operator). From this perspective, objective reduction would be regarded as a secondary, possibly stochastic, process which modifies unitary evolution during isolated intervals in which measurement takes place. In this paper, we propose an opposing theory, consistent with the speculation (Penrose, 1994)  that non-computability may be a fundamental feature of quantum  theory. Here we show explicitly, for two-state quantum systems, that the conventional  complex Hilbert space formulation can be derived from a generically non-computable real-number description of the quantum state, in which physically-realisable states are themselves computable.  From this perspective, measurement outcome is simply and deterministically defined. Complex numbers are nowhere introduced axiomatically into this  theory. 

As is well known, the Riemann sphere plays a fundamental role in the complex-number description of two-state quantum systems. In section \ref{sec:riemann} a  construction of the Argand plane and the Riemann sphere is put forward using as primary elements, not complex numbers, but certain divergent sequences $\mathcal{S}$ with bivalent elements. These sequences are related to  the binary expansion of generic real numbers $r$. Complex structure arises from  self-similar scaling properties defined from a set of bijections $i^{q}$, $q$ rational, acting on these sequences (for example, for all $\mathcal{S}$, $i^2(\mathcal{S})=-\mathcal{S}$). These induce the transformation $\tilde{r}=\tilde{i}^q(r)$ on the underlying reals. 

A description of the quantum state in terms of these reals is given in section \ref{sec:quantum}. A criterion for  measurement outcome is proposed based on whether or not $\tilde{r} \ge 1/2$. It is shown that for a generic point on $S$, this criterion cannot be determined by a Turing machine. However, it is shown that $\mathcal{S}$ can also represent a certain divergent sequence of recursive solutions to approximations of the measurement problem. It is demonstrated that this representation is equivalent to the conventional Dirac description of the quantum state vector as a complex linear superposition of measurement outcomes. 

In section \ref{sec:quantum}, it is also described how rotations of some preferred direction on $S$ (associated with oriented measurement) give rise to transformations of the sequences $\mathcal{S}$ which are equivalent to unitary transformations in the Heisenberg representation of quantum theory. Most importantly, a general rotation maps a computable real $r_p$ at a point $p$ to a non-computable real $r^{\prime}_p$. A simple trigonometric identity on the sphere is shown to give rise to the Heisenberg uncertainty principle, using the sequence representation of the quantum state. We note the special points $p_* \in S$ where  both $r_{p_*}$ and $r^{\prime}_{p_*}$ are computable, and from these doubly-computable points derive the observed quantum correlations for entangled particle states.
 
In section \ref{sec:euler}, it is suggested that the evolution of the real-number representation of the quantum state is given, in differential equation form, by the Euler equation. There are three reasons for this. Firstly, in view of  the success of geometric thinking in defining the fundamental laws of physics, it can be noted that smooth Euler flows on some compact Riemannian manifold $\mathcal{M}$ are geodesics on the Lie group of volume-preserving diffeomorphisms of $\mathcal{M}$ (Arnold and Khesin, 1998). Secondly, and most importantly, whilst the Euler equation is deterministic, scaling arguments (based on the self-similar energy cascade) suggest that, in finite time, solutions can be sensitive to indefinitely small-scale perturbations to the initial state, ie are non-computable.  Indeed rigorous non-computable solutions of the Euler equation have been recently found  (Scheffer, 1993; Shnirelman, 1997). Thirdly, through a non-linear Cole-Hopf transformation, there are some intriguing connections between the linear Schr\"{o}dinger equation, and non-linear Euler-type equations. It is shown how the sequences described in section  \ref{sec:riemann} can be associated with sequences of solutions to Galerkin truncations (ie Turing machine representations) of the underlying Euler equation  (over the set of all truncation wavenumbers). 

Finally, in section \ref{sec:ontology}, we explore the ontological implications of the isomorphic representations of the quantum state as a real number and as a sequence of recursive solutions to the measurement problem. For a given reference direction on $S$, we associate (the countable set of ) points $p \in S$ which have computable reals $r_p$ with physically-realisable states. The remaining (the uncountable set of) points $p^{\prime} \in S$ have non-computable reals $r_{p^{\prime}}$. A rotation of the reference direction in general maps a computable state to a non-computable state, and a physically-realisable state to a counterfactual state. We discuss the ambiguity in the notion of mathematical existence for these non-computable states; this ambiguity implies that the model is not constrained by Bell's inequality.

It is argued that in this approach to the formulation of quantum theory, physically-realisable quantum states have an objective reality (the latter word used in both its senses), and that their measurement requires no dice.  Some remarks on other specific quantum paradoxes (eg wave-particle duality, non-locality   and the many-worlds interpretation) are made, based on the properties of the real-number representation of the quantum state.  

\section {Non-Computable Reals and The Riemann Sphere}
\label{sec:riemann}

In this section we consider a 2-sphere $S$ of unit radius, whose points represent certain unending  sequences of `$1$'s and `$-1$'s. The relative fraction of `$1$'s in the sequence represented by some point $p$, depends on $p$'s latitude on $S$; the north and south poles represent the constant sequences $\{1,1,1,...\}$ and $\{-1,-1,-1...\}$ respectively, whilst equatorial points represent sequences with equal numbers of `$1$'s and `$-1$'s. We derive complex structure from a self-similarity property associated with a set of operators acting on these sequences. There is a fundamental isomorphic duality in the interpretation of these sequences: as the binary expansion of some generically non-computable real-number $0\le r\le1$, or as the sequence of outcomes  to recursive approximations to the `measurement problem': is $r\ge 1/2$ or $< 1/2$?  This duality is explained below.

\subsection{Points on the Equator: the Argand Plane}
\label{sec:argand}

Consider a real number $0<r<1$. For almost all $r$, the binary representation of $r$ contains, over any sufficiently long segment, as many `0's as `1's (eg Hardy and Wright, 1938). Let $\mathcal{S}=\{a_1, a_2, a_3, ....\}$ where $a_n \in \{1, -1\}$ denote the sequence obtained by replacing each occurrence of `0' in the binary representation of such a generic $r$, with `-1'. Hence the mean and variance, 
\begin{eqnarray} 
\mu(\mathcal{S})=\frac{1}{N}\sum_{i=1}^N a_i \rightarrow 0 \nonumber \\
\sigma^2(\mathcal{S})=\frac{1}{N} \sum_{i=1}^N (a_i-\mu)^2 \rightarrow 1 
\end{eqnarray}
as $N \rightarrow \infty$. Again, if $r$ is generic, then the real number constructed by taking all the odd elements, or all the even elements in the binary representation of $r$, will also be generic, ie will also have as many $1$s as $0$s. Indeed this is true for the real number constructed from any sufficiently long  nameable subset of elements of the binary representation of $r$. 

Let us assign some particular such $\mathcal{S}$ (and hence $r$) to a reference point $p_0$ on the equator of $S$, and define  this point to have a longitude $\lambda=0$, and latitude $\theta=0$. Now if $\mathcal{S}=\{a_1, a_2, a_3,.....\}$ is such a sequence with mean $\mu=0$ then so is $-\mathcal{S}= \{-a_1, -a_2, -a_3.....\}$.We associate $-\mathcal{S}$ with the point on the equator at longitude $\pi$,  antipodal to $p_0$ (see Fig 1). Clearly the real number associated this antipodal point is $1-r$.

Consider now the operator $i$ which is defined in two steps: first multiply the  second member of each pair of elements of $\mathcal{S}$ by $-1$, and then swap the order of the elements within each pair. That is
\be
\label{eq:i}
\mathcal{S}^{\prime}=i(\mathcal{S})=\{-a_2, a_1, -a_4, a_3, -a_6, a_5,.....\}
\ee

Since the subsequences of all even members and all odd members, have zero mean, then $\mu(\mathcal{S}^{\prime})=\mu(-\mathcal{S}^{\prime})=0$. We let $\mathcal{S}^{\prime}$ and $-\mathcal{S}^{\prime}$ be represented by points on the equator of $S$ at longitudes $\pi/2$ and $3 \pi/2$ respectively (see Fig 1). The operator $i$ induces the  transformation  $\tilde{i}(r)$ on $r$. The binary representation of $\tilde{i}(r)$ is given by equation  \ref{eq:i} with the replacement $-1 \rightarrow 0$. 

It is immediate from equation \ref{eq:i} that
\be
i^2(\mathcal{S})=-\mathcal{S}
\ee
where $i^2 \equiv i*i$. Moreover, since $0 \le\ \tilde{i}(r) \le 1$ is also a generic real number then we can similarly operate on $\mathcal{S}^{\prime}$ with $i$ and $i^2$, giving
\begin{eqnarray}
i^2(\mathcal{S}^{\prime}) &=& -\mathcal{S}^{\prime}\nonumber \\
i(-\mathcal{S}^{\prime}) &=& \mathcal{S}
\end{eqnarray}
We can generalise $i$ to define a set of operators represented by fractional powers $i^{1/{2^n}}$ of $i$. This representation is justified by the property, demonstrated below, that 
\be
\label{eq:power}
i^{1/{2^n}} i^{1/{2^n}}(\mathcal{S})=i^{1/{2^{n-1}}}(\mathcal{S})
\ee
Let us start with $n=1$. Just as $i$ ($n=0$) was defined in 2 steps on pairs of elements, so $i^{1/2}$ is defined in 3 steps on quadruplets of elements. In the first step, the last element of each quadruplet of elements is multiplied by $-1$, ie
\be
\{a_1, a_2, a_3, a_4, ..\} \rightarrow \{a_1,a_2,a_3-a_4, ...\}
\ee
In the second step, the elements within the last pair of each quadruplet are swapped, ie
\be
\{a_1,a_2,a_3,-a_4..\} \rightarrow \{a_1, a_2, -a_4, a_3..\}
\ee
Finally, the two pairs of elements in each quadruplet are swapped pairwise, ie
\be
\{a_1,a_2,-a_4,a_3..\}\rightarrow \{-a_4, a_3, a_1,a_2..\}
\ee
Putting this together, we have, for the first two quaduplets, 
\be
i^{1/2}(\mathcal{S})=\{-a_4, a_3, a_1,a_2,-a_8,a_7,a_5,a_6...\}
\ee

Again $i^{1/2}$ induces the transformation $\tilde{i}^{1/2}(r)$ on $r$. Because nameable subsequences of $\mathcal{S}$ have zero mean, then $\mu(i^{1/2}(\mathcal{S}))=0$ and $\tilde{i}^{1/2}(r)$ is a generic real number. We represent $i^{1/2}(\mathcal{S})$ at the point on the equator at longitude $\pi/4$ (see Fig 1). As before, we can operate on $i^{1/2}(\mathcal{S})$ with $i$ and $i^{1/2}$ yielding, for example
\begin{eqnarray}
i^{1/2}i^{1/2}(\mathcal{S}) &=&\{-a_2,a_1,-a_4,a_3,-a_6,a_5,-a_8,a_7\}=i(\mathcal{S})\nonumber \\
i^{1/2}*i(\mathcal{S}) &=& i*i^{1/2}(\mathcal{S})
\end{eqnarray}
 
More generally, it is possible to define $i^{1/{2^n}}$ by a self-similar rule operating on such  sequences. (Self-similar scaling is a fundamental property of non-computable solutions of the Euler equation put forward in section \ref{sec:euler}.) More specifically we define $i^{1/{2^n}}$ by an $(n+2)$-step operation on consecutive non-overlapping $2^{n+1}$-tuplets of $\mathcal{S}$. Each $2^{n+1}$-tuplet is split into $n+1$ subsequences of length $2^n, 2^{n-1}, 2^{n- 2},...,2^1,2^0+1$ (these numbers sum together to make $2^{n+1}$). The procedure on  each $2^{n+1}$-tuplet is as follows. Multiply the last element of the last pair by $-1$ (first step). Then swap the elements within this last pair of elements (second step). Then swap the two adjacent pairs of elements within the last quadruplet of elements (third step). Then swap the two adjacent quadruples of elements within the last octuplet of elements (fourth step). The (i+2)th step requires swapping the two adjacent $2^i$-tuplets within the last $2^{i+1}$-tuplet. The last step occurs when $i=n$.  Specifically, with
\be
\mathcal{S}=\{a_1,....,a_{2^{n+1}}, ...\}
\ee
then $i^{1/{2^n}}(\mathcal{S})$ is given by 
\be
\label{eq:sequence}
i^{1/{2^n}}(\mathcal{S})=\{\beta_1,.....,\beta_{n+1}...\} 
\ee
where each $\beta_i$ is a subsequence of $\mathcal{S}$ given by
\begin{eqnarray}
\label{eq:sequence1}
\beta_{n+1}&=&\{a_1, a_2,....a_{2^n}\} \nonumber \\
\beta_n &=&\{a_{2^n+1}, a_{2^n+2},.....,a_{2^n+2^{n-1}}\} \nonumber \\
\beta_{n-1}&=&\{a_{2^n+2^{n-1}+1}, a_{2^n+2^{n-1}+2},.....a_{2^n+2^{n-1}+2^{n-2}} \}\nonumber\\
.\nonumber\\
.\nonumber\\
\beta_2&=&\{a_{2^n+2^{n-1}+...+2^2+1}, a_{2^n+2^{n-1}+..2^2+2^1}\}\nonumber\\
\beta_1&=&\{-a_{2^n+2^{n-1}+..+2^0+1}, a_{2^n+2^{n-1}+..2^0}\}
\end{eqnarray}
The last two subsequences can equivalently be written as
\begin{eqnarray}
\label{eq:sequence2}
\beta_2=\{a_{2^{n+1}-3},a_ {2^{n+1}-2}\}\nonumber\\
\beta_1=\{-a_{2^{n+1}}, a_{2^{n+1}-1}\}
\end{eqnarray}
This whole procedure is applied separately to each consecutive $2^{n+1}$-tuplet. 

To show equation \ref{eq:power} we use the self-similarity of the operational procedure described above. Operate on $\{\beta_1, ... \beta_n, \beta_{n+1}..\}$
with $i^{1/2^n}$. For the first $2^{n+1}$-tuplet, the first $(n+1)$ steps of the operation are applied to elements of the subsequence $\beta_{n+1}=\{a_1, a_2...a_{2^n}\}$ (as detailed above). By definition this gives the first $2^n$ elements of $i^{2^{n- 1}}(\mathcal{S})$. The final $n+2$nd step involves swapping this subsequence with the subsequence $\{\beta_1...\beta_n\}$. The subsequence $\{\beta_1...\beta_n\}$ gives the next $2^n$ elements of $i^{2^{n-1}}(\mathcal{S})$. QED.

On this basis, we represent the sequence $i^{1/2^n}(\mathcal{S})$ by a point on the equator of $S$ with longitude $\pi/2^{n+1}$ radians. Moreover,  by using product operations such as
\be
\label{eq:times}
i*i^{1/{2^n}}(\mathcal{S}) = \underbrace{i^{1/{2^n}}*i^{1/{2^n}}....}_{2^n+1}=i^{1+1/{2^n}} (\mathcal{S})
\ee
we can define constructively any sequence $i^q(\mathcal{S})$, $q$ rational and associate it with the point on the equator of $S$ at longitude $\lambda=\pi q/2$. Hence, if $i^{q_1}(\mathcal{S})$ and $i^{q_2}(\mathcal{S})$ are sequences at points on the equator with longitudes $\lambda_1=\pi q_1/2$ and $\lambda_2=\pi q_2/2$, then $i^{q_1}*i^{q_2}(\mathcal{S})$ is a sequence at the point with longitude $\lambda_1+\lambda_2$. 

Since every sequence so generated has mean equal to zero (ie the real $\tilde{i}^{q}(r)$ is generic for all $q$), then the properties of the operator $i^{q}$ apply when operating on any of these sequences. In this sense, we can suppress the argument $\mathcal{S}$, and write, for example
\be
i^2=-1
\ee
Essentially, we have emulated complex multiplication on the unit circle of the equatorial plane of $S$, over the set of points whose longitude is a rational multiple of $\pi$ (the `rational' points), without having ever introduced complex numbers!

Having defined constructively, sequences (and associated reals) for the `rational' points on the equator of $S$, let us now consider the extension to  points on $S$ whose longitude is not a rational fraction of $\pi$ (the `irrational' points). Since $2^{\aleph_0}$ is the cardinality of the continuum, then by letting $n \rightarrow \infty$ in equations \ref{eq:sequence} and \ref{eq:sequence1}, we could non-constructively associate any point on the equator of $S$ whose longitude $\lambda$ is an irrational multiple of $\pi$, with the sequence $i^{2 \lambda/\pi}(\mathcal{S})$ of $1$s and $-1$s. This sequence would in turn represent some  real number $\tilde{i}^{2\lambda/\pi}(r)$.

However, this real number is non-computable. Whilst the points on $S$ representing $\mathcal{S}$ and $i^{1/2^n}(\mathcal{S})$ become  more and more adjacent the larger is $n$, there is no requirement that the corresponding numbers $r$ and $\tilde{i}^{1/2^n}(r)$ converge. For example, the $2^{n+1}$th element of $\mathcal{S}$ is $a_{2^{n+1}}$. With $-1 \rightarrow 0$ this defines the $2^{n+1}$th binary digit of the corresponding $r$. Hence for large $n$, $a_{2^{n+1}}$ makes a tiny contribution to $r$. On the other hand, from equations \ref{eq:sequence} and \ref{eq:sequence1} it can be seen that the first element of $i^{1/2^n}(\mathcal{S})$ (which determines whether or not on $\tilde{i}^{1/2^n}(r) \ge 1/2$) is $-a_{2^{n+1}}$, and thefore makes a big contribution to the value of $\tilde{i}^{1/2^n}(r)$. Since almost all  points $p$ on the equator are `irrational', then determining the value of $r_p$ (eg whether it is $\ge 1/2$ or $<1/2$) associated with some arbitrary point $p$ on the equator is generically non-computable. 

To make this point more explicitly, consider a Cauchy sequence of points $\{p_1,p_2, p_3....\}$ on the equator of $S$, which converges to some generic point $p$, such that each point $p_n$ has a longitude $\pi q_n/2$, $q$ rational. For each $p_n$ we can estimate (computably)  $i^{q_n}(\mathcal{S})$ and hence $\tilde{i}^{q_n}(r)$. However, no matter how many values $\tilde{i}^{q_n}$ ($n=1,2,...$) are calculated, we are no closer to knowing even the first digit in the binary expansion of $r_p$. This raises the question: do the real numbers $\tilde{i}^{2\lambda/\pi}(r)$, $\lambda$ an irrational multiple of $\pi$, ` really' exist? This matter is discussed in section \ref{sec:ontology}.

\begin{figure}
\epsfxsize=10cm
\centerline{\epsffile{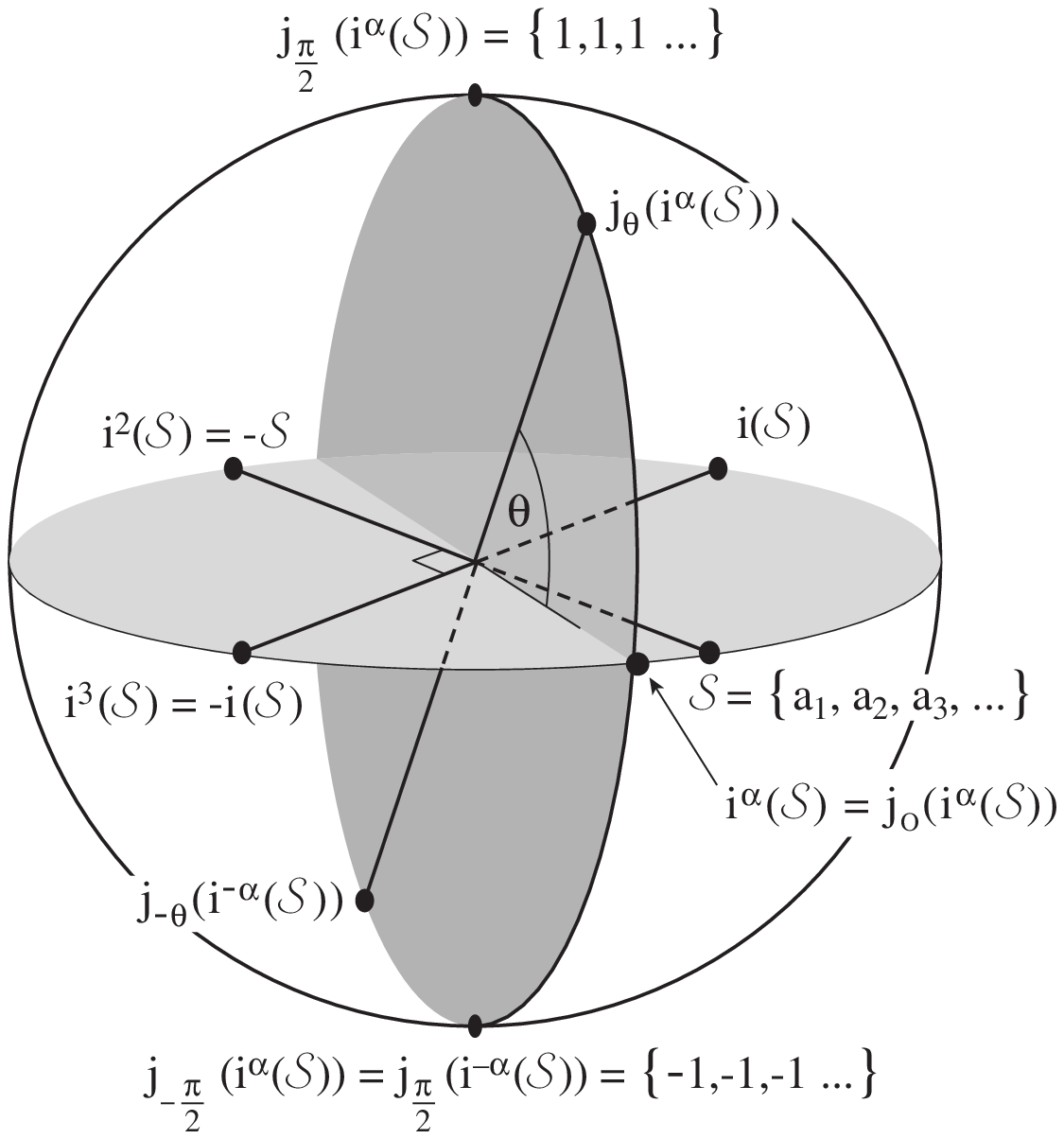}}
\caption{\textsf{A realisation of the Riemann sphere $S$ based on a reference sequence $\mathcal{S} \equiv \{a_1, a_2, a_3.....\}$ whose elements $a_n \in \{1, -1\}$ are related to the binary expansion of some generic real number $0 \le r \le 1$. The operators $i^q$, $q$ rational,  acting on $S$ have certain self-similar properties described in the text which give rise to a complex structure on the sequences. These operators induces transformations $\tilde{i}^q(r)$ of $r$. The operator $j_{\theta}$ maps sequences at equatorial points to sequences at other points on $S$. The construction is the basis of a demonstration that two-state quantum theory is obtainable from a generically non-computable real-number description of the quantum state and a simple deterministic criterion for measurement outcome.}}
\label{fig:Fig1}
\end{figure}

We conclude this section by noting that the method above can be generalised to define complex numbers on the whole equatorial plane (ie giving the Argand plane). Let 
\begin{eqnarray}
\label{eq:argand}
\mathcal{S}_{\alpha}& \equiv&  \{\alpha a_1,\alpha  a_2.....\}\nonumber \\
& \equiv& \alpha \{a_1, a_2...,a_i,..\}\nonumber \\
& \equiv & \alpha \mathcal{S}
\end{eqnarray}
where  $\alpha$ is an arbitrary real number. With $p$ representing $\mathcal{S}$, the sequence $\mathcal{S}_{\alpha}$ is represented by a point on the line in the equatorial plane joining the centre of $S$ with $p$, a distance $\alpha$ from the the centre. In terms of this, an operation `$\oplus$' can be defined by
\be
\label{eq:oplus}
\mathcal{S}_{\cos \lambda} \oplus (i( \mathcal{S}))_{\sin \lambda}=i^{2 \lambda/\pi}(\mathcal{S})
\ee
which parallels ordinary complex addition. One could think of `$\oplus$' as defining an `intertwining' of the elements of $\mathcal{S}$ with those of $i(\mathcal{S})$, with $\cos\lambda$ defining some weighting or preference towards the elements of one sequence over the other. (The operator $\oplus$ plays no essential role in the real-number representation of the quantum state, other than to relate $\mathcal{S}$ to the standard Dirac representation of the quantum state vector. 

\subsection{Points on a Meridian of the Riemann Sphere}
\label{sec:meridian}

Given a sequence $\mathcal{S}(0, \lambda)=\{c_1,c_2,c_3....\}$ associated (via the construction in section \ref{sec:argand}) with some point $p(0, \lambda)$ on the equator of $S$, we now define the sequences 
\be
\mathcal{S}(\theta, \lambda)=\{d_1, d_2, d_3,.....\}.
\ee
associated with points $p(\theta, \lambda)$ on the meridan through $p(0, \lambda)$. 

This is done in two steps. First $\mathcal{S}(0, \lambda)$ is mapped to points on the meridian with latitude $\theta=\pi q/2$, $q$ rational, using the operator $i^q$ as defined in the previous section. Suppose the resulting sequence is $\{c^{\prime}_1, ....c^{\prime}_n, c^{\prime}_{n+1}, c^{\prime}_{n+2}......\}$. Now let $0 \le r^{\prime}_n \le 1$ denote the real number associated with the sequence $\{ c^{\prime}_n, c^{\prime}_{n+1}, c^{\prime}_{n+2}......\}$ (where, as before, `$-1$' is replaced with `$0$' wherever it occurs in the sequence) ie starting with the $n$th element $c^{\prime}_n$. The definition of $d_n$ is as follows: if $r^{\prime}_n \ge (1-\sin\theta)/2$, then let $d_n=1$; otherwise let $d_n=-1$. Based on these two sub-operations, we define the compound operator $j_{\theta}$, so that $j_{\theta}(S(0,\lambda)=S(\theta, \lambda)$ (see Fig 1). 

The following can be readily deduced (see Fig 1):
\begin{eqnarray}
j_0(\mathcal{S}(0,\lambda))&=&\mathcal{S}(0,\lambda)\nonumber\\
j_{\pi/2}(\mathcal{S}(0,\lambda)) &=& \{1,1,1,....\}\nonumber\\
j_{-\pi/2}(\mathcal{S}(0,\lambda))&=&\{-1,-1,-1....\}\nonumber\\
\mathcal{S}(\theta,\lambda) &=&-\mathcal{S}(-\theta, -\lambda).
\end{eqnarray}
Moreover, since $0\le r^{\prime}_n\le 1$ with uniform probability in $[0,1]$, the probability that $r^{\prime}_n < (1-\sin \theta)/2$ (ie that $d_n=-1$) is equal to $(1-\sin \theta)/2$. Hence $\mu(j_{\theta}(\mathcal{S}(0,\lambda)))=\sin \theta = \cos \tilde{\theta}$ where $\tilde{\theta}= \pi/2-\theta$ is co-latitude. Since the mean value of the set of elements in $j_{\theta}(\mathcal{S}^{\prime})$ is $\cos\tilde{\theta}$, it is easily shown that the standard deviation $\sigma(j_{\theta}(\mathcal{S}(0,\lambda)))$ is necessarily equal to $|\sin\tilde{\theta}|$. This standard deviation defines an intrinsic `uncertainty' in the value of elements of  $j_{\theta}(\mathcal{S}(0,\lambda))$ (see section \ref{sec:uncertainty}). As in section \ref{sec:argand}, the state $r_p$ at $p(\theta,\lambda)$ has been defined constructively for a countable number of rational points. The other points on $S$ are, as before, associated with  non-computable reals (defined by the extension  $i^{\alpha}$ for irrational $\alpha$). 

As with equation \ref{eq:oplus} an operator `$\oplus$' can also be defined so that, using equation \ref{eq:argand}
\be
\label{eq:add}
\mathcal{S}_{\cos\theta}(0,\lambda)\oplus\mathcal{S}_{\sin\theta}(\pi/2, \lambda)=\mathcal{S}(\theta, \lambda)
\ee
Again, this could be thought of as defining a suitably-weighted intertwining of the elements of $\mathcal{S}(0,\lambda)$ with the elements of $\mathcal{S}(\pi/2, \lambda)$. With $\mathcal{S}(\pi/2, \lambda)= \{1,1,1,....\}= \mathbf{1}$ and $\mathcal{S} (0,\lambda)=i^{2\lambda/\pi}(\mathcal{S})$, then equation \ref{eq:add} can be written 
\be
\label{eq:add1}
\mathcal{S}(\theta, \lambda)=\cos \tilde{\theta} \mathbf{1} \oplus \sin \tilde{\theta} i^{2\lambda/\pi}(\mathcal{S}).
\ee
It can be noted that $\oplus$ is not needed in the development of the current theory, except to compare with the conventional Dirac representation of the quantum state. 

\section{A Real-Number Description of the Quantum State}
\label{sec:quantum}

Based on the constructions in the previous sections, we put forward a local deterministic real- number description of the quantum state, associated with two observable states $\mathbf{1}$ and $\mathbf{-1}$. The configuration space of the quantum state is the 2-sphere $S$. The set of points on $S$ can be identified with the set of directions on the celestial sphere at some point in physical space.

A quantum state is defined with respect to some direction on the celestial sphere; this direction defines an orientation of some measuring system. Hence, we cover $S$ with  latitude/longitude coordinates $(\theta, \lambda)$, such that the measuring system is oriented in the direction $\theta=\pi/2$ on the celestial sphere.  Relative to this orientation, the quantum state at each point $p \in S$ is given by a real number $0\le r_p\le1$. We assert (by axiom) the existence of a single quantum state (represented by some $r$) at some reference point on $S$ with coordinates $(0,0)$.

The states $r_p$ associated with the points $p(\theta, \lambda) \in S$ with coordinates are determined by taking the binary expansion of $r$, replacing each $0$ with a $-1$, and applying the transforms $i^{2 \lambda/\pi}$ and $j_{\theta}$ to the resulting sequence $\mathcal{S}$, (giving $\mathcal{S}(\theta, \lambda)$) finally transforming back ($-1 \rightarrow 0$) to give the binary expansion of $r_p$. If $\lambda$, $\theta$ is a rational multiple of $\pi$, $r_p$ is a computable state. 

\subsection{Criterion for Measurement Outcome - Duality in State Representation}
\label{sec:measurement}

Our criterion for the outcome of an oriented measurement is very simply stated: if measurement takes place, then the observable is $\mathbf{1}$ if $r \ge 1/2$, and $ \mathbf{-1}$ if $r <1/2$. (We do not say in this paper what constitutes the process of measurement. Following Penrose, 1994, the author is attracted to the idea of some relatively- simple gravitationally-induced instability. In such circumstances,  the direction of instability - towards outcome $\mathbf{1}$ or $\mathbf{-1}$ - is trivially determined by the value of $r_p$)  As discussed in section \ref{sec:riemann}, for a given measurement orientation, there is a countable set of points where the quantum state (as stated above) is a computable real number. However, for almost all $p \in S$, the quantum state is a non-computable real number. In essence, the set of points $p$ where the corresponding real $r_p \ge 1/2$,  is infinitely intertwined with the set of points $p$ where $r_p < 1/2$ (reminiscent of non-linear riddled-basin dynamics: Sommerer and Ott, 1993, Palmer, 1995, Nicholis et al, 2001). Hence, for an arbitrary $p$, the measurement problem cannot be determined by a Turing machine. As discussed below, we necessarily need to consider these non-computable states when considering the effect on the quantum state of a change in the orientation of our measuring system with respect to the celestial sphere. 

For a general point $p \in S$, there are (many) Cauchy sequences of points  $\{p_1, p_2...\}$,  $p_n \in S$ which converge to $p$ and where the outcome of measurement ($\mathbf{1}$ or $\mathbf{-1}$) associated with any of the $p_n$ can be computably estimated, given the reference state $r$. Hence for each Cauchy sequence of points converging on $p$, there is a divergent sequence of computable solutions ($\mathbf{1}$ or $\mathbf{-1}$) to the measurement problem. Since the set of points on $S$ where $r \ge 1/2$, is infinitely intertwined with the set of points where $r < 1/2$, then in fact we can interpret $\mathcal{S}(\theta, \lambda)$ as defining one such sequence of computable solutions associated with a particular Cauchy sequence of  points on $S$. Of course, just as $r_p$ is non-computable, so the latitude/longitude coordinates of the Cauchy sequence of points which generate $\mathcal{S}$ are also non-computable. 

Moreoever, $\mathcal{S}(\theta, \lambda)$ can also be represented by conventional Dirac notation. To see this, refer back equation \ref{eq:add1} and let the constant sequences $\{1,1,1...\}$ and $\{-1,-1,-1...\}$ correspond to the pure states  $|\mathbf{1}\rangle$ and $|\mathbf{-1}\rangle$ respectively, and let $\mathcal{S}(\theta, \lambda)$ correspond to the mixed state
\be 
\label{eq:dirac}
|\psi\rangle= \cos \frac{\tilde{\theta}}{2} |\mathbf{1}\rangle + e^{i\lambda} \sin \frac{\tilde{\theta}}{2} |\mathbf{- 1}\rangle
\ee
where $\tilde{\theta}$ is co-latitude, and $i$ is now, of course, $\sqrt{-1}$. The factor of $1/2$ in co-latitude between equations \ref{eq:add1} and \ref{eq:dirac} is easily explained. The `$+$' sign in equation \ref{eq:dirac} can be thought of (in the present context) as defining a generically  non-computable merging of the elements of $\{1,1,1....\}$ with the elements of $\{-1,-1,-1....\}$. On the other hand, the `$\oplus$' sign in equation \ref{eq:add1} can be thought of as a merging of $\{1,1,1....\}$ with the sequence $\mathcal{S}(0, \lambda)$ (which has equal numbers of `$1$'s and `$-1$'s).

Hence there is a fundamentally-important isomorphism between the representation of a general quantum state as a non-computable real, as a divergent sequence of solutions of recursive approximations to the measurement problem, and as a complex linear combination of measurement outcomes.

\subsection{Unitary Evolution}
\label{sec:unitary}

In the next four subsections we consider how the state $r_p$ transforms under a coordinate transformation, associated with a rotation of the orientation of the measuring device with respect to the celestial sphere.  Let us start with the simple case of  a constant angular rotation $\omega$ about the $\theta=\pi/2$ axis. If $t$ is time, then from section \ref{sec:argand}, this  induces the change
\be
\label{eq:time}
\mathcal{S}(t)=i^{\frac{2 \omega}{\pi}(t-t_0)} \mathcal{S}(t_0)
\ee
to the sequence $\mathcal{S}(t_0)$, associated with some point $p \in S$. With the identification of $\mathcal{S}$ with $|\psi\rangle$ (cf equation \ref{eq:dirac}), this is equivalent to the first integral
\be
|\psi(t) \rangle = e^{iH(t-t_0)/\hbar}|\psi(0)\rangle
\ee
of the Schr\"{o}dinger equation for stationary states, with $\omega=H/\hbar$.  The effect on $\mathcal{S}$ of a rotation of the zero of longitude about the north pole, is therefore equivalent to a multiplication of $| \psi \rangle$ by a complex phase factor. 

Let us write
\be
\label{eq:sch}
\mathcal{S}(t)=U(t,t_0)\mathcal{S}(t_0)
\ee
 to denote a general transformation of the sequence representation of the state vector, under such a rotation. The corresponding real number representation of the state vector can be written as
\be
\label{eq:eul}
\tilde{r}(t)=\tilde{U}(t, t_0) (r(t_0))
\ee
If equation \ref{eq:sch} is a first-integral of the complex Schr\"{o}dinger equation, of  what differential equation is \ref{eq:eul} the first-integral? Certainly equation \ref{eq:eul} will not have the simple analytic properties of equation \ref{eq:sch}. We return to this question in section \ref{sec:euler}.

\subsection{Non-Computability and Counterfactuality}
\label{sec:counterfactuality}

Let us now consider the ontologically much more important situation where the orientation of the measuring system rotates from the north pole to some arbitrary point with coordinates $(\theta_0, \lambda_0)$. Let  $r_p$ denote the state at some general point $p \in S$ when the reference direction is $\theta=\pi/2$, and let $r^{\prime}_p$ denote the state at $p$ relative to the $(\theta_0, \lambda_0)$ direction. If $r_p$ and $r^{\prime}_p$ were both unambiguously well defined, this would imply that simultaneous measurement relative to both directions would have unambiguously well-defined outcomes.

As is well known (see also discussion in section \ref{sec:ontology}), if this were to be the case, then the model would be in the form of a local hidden-variable model, and would therefore necessarily be constrained by Bell's inequality, and be inconsistent with observations. However, as we now discuss, if $r_p$ is a computable real, then in general, $r^{\prime}_p$ is a non-computable real whose existence is mathematically ambiguous.   

To see this let us computably generate $r_p$ on a finite number $N$ of  meridians $\lambda=2 \pi n/N$, $n=1,2,...N$ and at a finite number $(2N+1)$ of latitude circles $\theta=\pm \pi m/2N$, $m=0,1,...N$, using, as before,  the operators $i^q$ and $j_{\theta}$ and the reference real $r$. Notice that the density of points with computably-defined states increases towards the poles (defining the reference direction). 

Now let $(\theta^{\prime}$, $\lambda^{\prime})$ denote latitude/longitude coordinates with respect to the transformed pole at $(\theta_0, \lambda_0)$. Then, as before, we can also computably generate a set of reals on the $N$ meridians $\lambda^{\prime}= 2 \pi n^{\prime}/N$ and at a finite number $(2N+1)$ of latitude circles $\theta^{\prime}=\pm \pi m^{\prime}/2N$, $m^{\prime}=0,1,...N$. Now in general, for arbitrary $(\theta_0, \lambda_0)$, the two sets of points where the real-number state is computably defined, need not intersect, and indeed in general will not intersect.  For example, if $\lambda_0=0$ then from elementary spherical trigonometry
\be
\label{eq:rotation}
\sin \theta^{\prime}=\cos(\pi m/2N) \cos (2 \pi n/N) \cos \theta_0 \pm \sin (\pi m/2N) \sin \theta_0.
\ee
In general, this implies that $\theta^{\prime}$ is not of the form $\theta^{\prime}=\pm \pi m^{\prime}/2N$. In other words, under a general  rotation of the pole, none of the original $N$ computed states needs map to any of the $N$ computed states in the rotated coordinates. In fact, since countable points are a set of measure zero in the continuum, then,  as $N \rightarrow \infty$, for an arbitrary point $p \in S$, the transformation $r_p \mapsto r^{\prime}_p$ associated with a general rotation of the pole on $S$ appears to take a computable real to a non-computable real. 

Consider the a physically-realisable proposition $\mathcal{P}$: `The outcome of a measurement with respect to the orientation $\theta=\pi/2$ is $\mathbf{1}$', determined by whether the computable $r_p \ge 1/2$. Consider also the counterfactual statement $\mathcal{P}^{\prime}$: `If a measurement in the direction $(\theta_0, \lambda_0)$ were to have been made, a measurement in the direction $\theta=\pi/2$ having in fact been made, then a definite outcome, either $\mathbf{1}$ or $\mathbf{-1}$, would have occurred.' Then  $\mathcal{P}^{\prime}$  will in general be determined by whether the non-computable real $r^{\prime}_p$ is $\ge 1/2$ or not. Does $\mathcal{P}^{\prime}$ have a well-defined truth value? We discuss this in section \ref{sec:ontology}. 

\subsection{The Uncertainty Principle}
\label{sec:uncertainty}

As has been discussed, there is a fundamental duality in the representation of a quantum state at some general $p \in S$; as a non-computable real, or as a divergent sequence of computable solutions to the measurement problem, itself representable as a complex linear combination of measurement outcomes. The key linking these representations is the sequence $\mathcal{S}(\theta, \lambda)$. The mean value of the elements of $\mathcal{S}(\theta,\lambda)$ is
\be
\mu(\mathcal{S}(\theta, \lambda)) \equiv \mu_{\tilde{\theta}} = \cos \tilde{\theta}
\ee
where $\tilde{\theta}$ is co-latitude. Since the individual elements of $\mathcal{S}(\theta, \lambda)$ are all `$1$'s and `$-1$'s, then it is easily shown that the standard deviation of the elements is given by
\be
\sigma(\mathcal{S}(\theta, \lambda)) \equiv \sigma_{\tilde{\theta}}=|\sin\tilde{\theta}|.
\ee
This standard deviation represents an intrinsic and irreducible uncertainty in the elements of the set of recursive approximations to the measurement problem.

Let us consider a rotation of the orientation of the measuring device by (approximately) $\pi/2$ radians, mapping the north pole to the equator at a longitude of (approximately) $\pi/2$ radians. (We use the word `approximately' here to denote the fact that in the following analysis, it is unimportant whether or not the transformed pole is at a rational or irrational point.) Then using elementary spherical trigonometry,  a point $p$ with co-latitude $\cos\tilde{\theta}$ and longitude $\lambda$ will have a co-latitude $\tilde{\theta}^{\prime}$ with respect to the rotated pole given by
\be
\label{eq:heisenberg}
\sin \tilde{\theta} \sin\lambda = \cos \tilde{\theta}^{\prime}.
\ee
We can interpret this simple trigonometric identity in terms of the sequence representation of the quantum state. We already have $\sigma_{\tilde{\theta}}= |\sin \tilde{\theta}|$ and
$\mu_{\tilde{\theta}^{\prime}}= \cos \tilde{\theta}^{\prime}$. The quantity $|\sin\lambda|$ can be interpreted as the standard deviation of the sequence $\mathcal{S}(\theta, \lambda)$ when the north pole is rotated by (approximately) $\pi/2$ radians to the equator at longitude of (approximately) zero, so that $\lambda$ is the co-latitude with respect to this (second) transformed pole. Hence, equation \ref{eq:heisenberg} implies
\be
\label{eq:uncertainty}
\sigma_{\tilde{\theta}} \sigma_{\lambda} = |\mu_{\tilde{\theta}^{\prime}}| \ge \frac{1}{2} |\mu_{\tilde{\theta}^{\prime}}|.
\ee
This is essentially a version of the uncertainty principle associated with measurements in the three orthogonal directions of the Euclidean space which embeds $S$. Of course, we could represent these directions in terms of the Lie algebra of generators of  $SO(3)$, the generator representing the right hand side of the inequality being given by the Lie bracket (commutator) of the other two generators. This would bring equation \ref{eq:uncertainty} into the standard form for the uncertainty principle in quantum theory. However, we will not labour this point here. 

The physical origin of  `uncertainty' is the intrinsic ambiguity in the recursive representation of the non-computable real-number quantum state at some $p \in S$, for different measurement orientations. 

\subsection{Entanglement}
\label{sec:entanglement}

As discussed, a computable state $r_p$ maps to a non-computable state $r^{\prime}_p$ at some general point $p$ under a general rotation of the measurement direction. However, we consider here the special set $S_{p_*}$ which is a subset both of $S_p$, the points which have computable states with respect to the original measurement direction, and $S^{\prime}_p$, the points which have computable states with respect to the transformed measurement direction.

Since meridians converge onto the poles, the density of points with computable states increases without bound in the neighbourhood of the north and south poles in the original and transformed coordinates. Hence some $p_*$ will occur in the neighbourhoods of the poles. For these particular doubly-computable points, it is easy to compute the correlation in measured outcomes with respect to measurements in the two directions. Focus on a particular point $p_*$ near the original north pole. The co-latitude of $p_*$ relative to the transformed north pole is $\pi/2-\theta_0=\tilde{\theta}_0$. Now because $p_*$ is near the north pole, the probability that $r_{p_*} \ge 1/2$ (ie outcome is $\mathbf{1}$) is close to 1. Similarly the probability that $r^{\prime}_{p_*} \ge 1/2$ is (by section \ref{sec:meridian}) close to $(1+\cos\tilde{\theta}_0)/2$. Now suppose $p_*$ is in the neighbourhood of the transformed north pole. Then the probability that $r_{p_*} \ge 1/2$ is $(1+\cos\tilde{\theta}_0)/2$, and the probability that $r^{\prime}_{p_*} \ge 1/2$ is close to 1. In both cases, the probability that both $r_{p_*} \ge 1/2$ and $r^{\prime}_{p_*} \ge 1/2$ is equal to $(1+\cos\tilde{\theta}_0)/2=\cos^2(\tilde{\theta}_0/2)$. Similar arguments apply to doubly- computable points $p_*$ chosen from the original south or transformed south poles: if $p_*$ is in the neighbourhood of the south pole, then $r_{p_*} < 1/2$ (ie outcome is $\mathbf{- 1}$) with probability close to $1$, and $r^{\prime}_{p_*} < 1/2$ with probability close to $(1+\cos\tilde{\theta}_0)/2$. Finally, if $p_*$ is in the neighbourhood of the south pole, then  $r_{p_*}<1/2$ with probability close to $1$, and $r^{\prime}_{p_*}\ge 1/2$ with probability close to $(1-\cos\tilde{\theta}_0)/2 = \sin^2(\tilde{\theta}_0/2)$.

Now $S_{p_*}$ is not a random subset of either $S_p$ or $S^{\prime}_p$ and as such the statistical properties of the states $r_{p_*}$ for some general doubly-computable point $p_* \in S$ do not necessarily inherit the statistical properties of either $r_p$ or $r^{\prime}_p$. (For example, consider the subset $S_1$ of $S_p$ where $p \in S_1$ iff $r_p \ge  1/2$. Clearly for some $p(\theta, \lambda) \in S_1$, the probability that $r_p < 1/2$ is not equal to $(1-\sin\theta)/2$ except at the pole $\theta=\pi/2$.) Now the actual values of $r_p$ and $r^{\prime}_p$ are determined by the reference reals $r$ and $r^{\prime}$. Up to now the results we have derived have been independent of the actual values of $r$ and $r^{\prime}$. We now assume that there are values for $r$ and $r^{\prime}$  such that the statistical properties of the states $r_{p_*}$ are independent of the location of $p_*$ on $S$ and therefore acquire the properties of $r_{p_*}$ for $p_*$ in the neighbourhood of the poles (so, for example, the probability that $r_{p_*} \ge 1/2$ and $r^{\prime}_{p_*} \ge 1/2$ is $(1+\cos\tilde{\theta_0})/2$ for all $p_* \in S$.  

As such, the proposed theory can describe the correct quantum correlations for entangled particle pairs. Let us describe the standard EPR experiment in the framework of the proposed theory. We imagine a source emitting entangled spin-1/2 particle pairs, and a pair of distant Stern-Gerlach devices measuring the spin of these particles with respect to two directions on the celestial sphere with relative orientation $\Delta \theta$. We represent the entangled quantum states as the reals $r_{p_*}$ and $r^{\prime}_{-p_*}=1-r^{\prime}_{p_*}$  at  antipodal points $p_*$ and $-p_*$ on $S$. Since the proposed measurements are physically realisable then we require these states to be doubly-computable real numbers. Let
\be
C(\Delta \theta)=\frac{1}{M}\sum_{n=1}^M (o_n o^{\prime}_n)
\ee
denote the correlation in spin outcomes $o_n= \pm1$ and $o^{\prime}_n=\pm 1$ associated with the two measurements. From the discussion above, we have (summing over the 4 pairs of outcomes $(o_n=1,o^{\prime}_n=-1;o_n=-1,o^{\prime}_n=1;o_n=1,o^{\prime}_n=1;
o_n=-1,o^{\prime}_n= -1)$, 
\begin{eqnarray}
\label{eq:correlation}
C(\Delta \theta) &=&- \frac{1}{4}[\cos^2(\Delta \theta/2)+\cos^2(\Delta \theta/2)+\sin^2(\Delta \theta/2)+\sin^2(\Delta \theta/2)] \nonumber \\
 &=&-\cos(\Delta\theta).
\end{eqnarray}
These outcomes are precisely what would have been obtained using the standard Dirac inner product $\langle \psi|\chi\rangle$ (cf equation \ref{eq:dirac}).
 
Let us now ask a crucial (counterfactual) question. Instead of the two actual directions of measurement, suppose  the spin of the second particle stream had been measured with respect to a third direction $\theta^{\prime\prime}, \lambda^{\prime\prime}$. Then, could we have obtained the correlation given by equation \ref{eq:correlation}? The answer is no, because the points $p_*$, which are doubly-computable with respect to the first and second measurement directions, will not be doubly-computable with respect to the first and third measurement directions (given the generic mapping of computable points to non-computable points, as discsussed in section \ref{sec:counterfactuality}). 

We refer to section \ref{sec:ontology} for further discussion of this aspect of entanglement and  its relation to non-locality theorems. 

\section{A Differential-Equation Approach to Real-Number Quantum State Evolution}
\label{sec:euler}

In section \ref{sec:unitary} we described a simple situation where the sequence representation of the quantum state evolved as an equivalent of a unitary transformation, and where equation \ref{eq:sch} represented a first integral of the Schr\"{o}dinger equation. We now ask what is the corresponding  differential equation that governs the evolution (whose first integral is given by equation \ref{eq:eul}) of the real-number representation of the quantum state. We argue in this section that there is strong evidence that this differential equation may be the familiar Euler equation for inviscid fluid flow. The argument is in two parts. Firstly it is shown that the Euler equation (which itself has a simple geometric formulation) appears to have the required non-computability properties. Secondly through the Cole-Hopf  transform, we shown that the non-linear Euler equation is transformable to the linear Schr\"{o}dinger equation.   

\subsection{The Euler Equation}
\label{sec:eulereq}

The profound beauty of the (diffeomorphism-invariant) theory of general relativity, the possible role that gravity may have in quantum state measurement(Penrose, 1994), and the discussion in the previous sections, motivates our search for a specific deterministic geometric non-computable differential equation for real-number quantum evolution. It is suggested that the Euler equation may form the basis of such a model. We also note that although the Euler equation is overtly reversible, its solutions may actually be irreversible. It has been suggested from cosmological considerations (Penrose, 1994) that the underlying quantum equations which govern the evolution of space-time may have a similar property. 

First we write the incompressible Euler equation in the conventional way in Euclidean space
\be
\label{eq:euler}
\frac{\partial u}{\partial t} +u \cdot \nabla u + \nabla p = 0 \\
\ee
where $u(x,t)$ is a 3-vector satisfying $\nabla \cdot u=0$. The scalar field $p$ is not independent, as can be seen by taking the divergence of equation \ref{eq:euler}. Letting $u^ \flat$ denote the 1-form associated with $u$, then the Euler equation 
\be
\label{eq:euler2}
\frac{\partial u}{\partial t}+ \mathcal{L}_u u^{ \flat} =dp^{\prime}
\ee
where $\mathcal{L}$ is the Lie derivative and $p^{\prime} = p-1/2 \mid u\mid^2$ can also applied to a general compact Riemannian manifold $\mathcal{M}$. Taking the exterior derivative of equation \ref{eq:euler2}, and letting $\omega=d u^{\flat}$ and $U$ denote (Newtonian) 4-velocity we have the simple geometric form
\be
\label{eq:euler3}
\mathcal{L}_U \omega = 0
\ee
The underlying geometric interpretation of the Euler equation can also be demonstrated by writing 
\be
\frac{\partial}{\partial t} \eta(t,x) = u(t, \eta(t,x))
\ee
where $\eta(x,t)$ denotes Lagrangian displacement with $\eta(0,x)=x$ for all $x \in \mathcal{M}$. Then the map $\eta_t: x \mapsto \eta(x,t)$ belongs to the group $\mathcal{D}_{\mu}(\mathcal{M})$ of volume preserving diffeomorphisms of $\mathcal{M}$. Arnold's theorem (eg Arnold and Khesin;1998) states that $u$ satisfies the Euler equations if and only if $\eta_t$ is a geodesic of the right-invariant $L^2$ metric on $\mathcal{D}_{\mu}$, defined by the kinetic energy integral over $\mathcal{M}$. 

Whether one views the Euler equation in terms of its differential equation form, as given in \ref{eq:euler}, \ref{eq:euler2}, \ref{eq:euler3} or the variational form of Arnold's theorem, the equation is essentially deterministic. In particular, there are no overtly stochastic elements in any of the defining equation forms. In this sense, given some exact initial state, the equations determine some precise future state (though see the comments at the end of this section). 

Despite determinism, it is known that finite-time solutions to the Euler equation do not have what is usually referred to in partial differential equation theory as a `uniqueness' property  (Scheffer, 1993, Shnirelman, 1997). This is a somewhat misleading word, at least in the present context, since the underlying equations of motion are deterministic and therefore unique for an \textit{exact} initial state. For reasons discussed below, this property is referred to here as  non-computability. The physical basis for non-computability lies in the well-known energy cascade associated with turbulent solutions to the Euler equation. Consider the famous self-similar Komolgorov (eg Frisch, 1995) scaling associated with  a 3-dimensional viscous turbulent fluid being forced (eg stirred) at some large scale. Energy associated with the forcing cascades to small scales and is ultimately dissipated at viscous scales. By simple scaling arguments, the  kinetic energy $E(k)$ per unit wavenumber $k$ in the inertial range between forcing and dissipation scales, varies  as $k^{-5/3}$. (The self-similarity of the Komolgorov cascade is reminiscent of  the self-similar properties of the operator $i^q$ in section \ref{sec:riemann}.) 

The predictability properties of such `5/3' turbulent solutions lie at the heart of non- computability constructions, as can be seen from the following scaling argument. Consider a time integration of the governing equations of a 3-dimensional fluid, with an initial state which is known perfectly at wavenumbers $\le k_r$ (within the inertial range), but is poorly known for wavenumbers $ > k_r$. Following Lorenz (1969),  assume that the time it takes for complete uncertainty at wavenumber $2k$ to strongly infect wavenumber $k$, is proportional to the `eddy turn-over time'  $\tau (k)=k^{-3/2} E(k)^{-1/2}$. The time $\Omega(N)$ taken for uncertainty to propagate from wavenumber $2^N k_L$ to wavenumber $k_L$ (where $k_L$ is a large-scale low wavenumber) is therefore given by 
\be
\label{eq:omega}
\Omega(N) \equiv \sum_{n=0}^{N-1} \tau(2^nk_L)
\ee
If $E(k) \sim k^{-5/3}$ and $\tau \sim k^{-2/3}$, then $\Omega(N)$ tends to a finite limit as $N \rightarrow \infty$, that is 
\be
\label{eq:omega1}
\Omega(\infty) \sim 2.7 \tau(k_L)
\ee

In the inviscid Euler limit, the inertial range will extend to arbitrarily small scales, and equation \ref{eq:omega1} implies that uncertainty in arbitrarily-small scales will imply, in finite time, uncertainty about  large scales. This loss of predictability is fundamentally greater  than that associated with chaotic ODEs. In the latter case, sequences of  finite time predictions will converge to some  true or exact state, if the corresponding Cauchy sequence of initial states converges to the true initial state. In the former case, weak convergence of the initial conditions to some exact initial state, does not guarantee convergence of the finite-time solutions to the `true' finite-time state. 

In fact, Shnirelman's (1999) construction is not based on the initial value problem in 3-dimensional Euler flow, but (for reasons of mathematical tractability) in 2-dimensional flow forced at small scales (for which the above scaling argument applies equally well; Kraichnan, 1967). Shnirelman's finds solutions $u_i$ with prescribed impulsive forcing $f_i$ which  converge weakly to zero as $i$ increases. The resulting $u_i$ converge to a (weak) solution $u$ of the homogeneous Euler equation which has the  property of being identically zero outside some finite time interval $\Delta T$, but not within this interval. The equations satisfied by weak solutions to the Euler problem cannot distinguish a forcing which is identically zero (so that $u=0$ in $\Delta T$), to  a forcing that is the limit of a sequence (of computable forces) which weakly converges to zero (so that $u \ne 0$ in $\Delta T$). Rigorous proofs of non-computability for the 3-dimensional Euler equation have not yet been given, and, although strongly supported by the Shnirelman construction, must nevertheless be considered putative at this stage.     

The reason that the phrase `non-computability' is used to describe this property is as follows. Suppose we project the initial state and the Euler equation onto an orthonormal basis (eg spherical harmonics, finite elements, wavelets etc) truncated to $n$ basis vectors, to produce an $n$-dimensional set of ODEs. Call this a Galerkin-$n$ truncation. (An example of the application of a hierarchical system of deterministic (hyperchaotic) ODEs to the quantum EPR problem is discussed in Duane, 2001).  Then non-computability implies that the sequence of solutions to the corresponding hierarchical set of Galerkin-$n$ truncated equations, will not converge as $n \rightarrow \infty$ (ie as more and more scales are introduced into the calculation). Since any Galerkin-$n$ integration is representable by a programmed Turing machine, then non-computability is equivalent to saying that the solution to the Euler integration cannot be represented by any finite number of programmed Turing machines. (It can be noted that the proof of existence of weak solutions to the Euler and Navier-Stokes equations depend on a projection onto such Galerkin bases.) Solutions to the Euler equation which do have the non-computability property cannot be smooth (ie classical). For example, given a smooth large-scale initial state, numerical integrations show that energy readily propagates downscale in the form of a front, suggesting (but not proving) the development of a singularity in the vorticity $\omega$ in finite time (eg Frisch, 1995). On the other hand, the regularity of the displacement and velocity fields associated with physically-meaningful weak solutions of the Euler equation are far from clear. The solutions to the Euler equation that correspond to the inviscid limit of the Kolmogorov cascade should exhibit finite energy dissipation at infinitesimal scales. As a result, one should expect physically appropriate solutions to the Euler equation to be irreversible even though the equations themselves are overtly reversible. In fact Shnirelma (2001) has recently constructed rigorous weak solutions of the Euler equation in which the kinetic energy is a decreasing function of time. It is presently unknown whether such types of solution have the non-computability property (A. Shnirelman, personal communication; 2000). Indeed, understanding the generic regularity properties of solutions to the Euler and Navier-Stokes equations remains the most celebrated of unsolved problems in the theory of partial differential equations.

\subsection{The Cole-Hopf Transform}
\label{sec:hopf}

In order to provide some further motivation for the use of the Euler equationas a model to describe the evolution of the real-number representation of the quantum state, consider the viscous Navier-Stokes equations
\be
\frac{\partial u}{\partial t} +u \cdot \nabla u + \nabla p = \nu \nabla^2 u \\
\ee
The Cole-Hopf  (eg Whitham, 1974) transformation defines a variable $\psi$ such that 
\be
u=-2\nu[\nabla \psi]/\psi
\ee
In terms of this transformation, the Euler equation can be put into the linear form
\be
\frac{\partial \psi}{\partial t}=\frac{p}{2 \nu}\psi+\nu \nabla^2\psi
\ee
With $p$ is considered fixed (the so-called Burgers equation), this is in the form of a linear Schr\"{o}dinger equation, but with $t \mapsto it$, and with additional dissipative forcing (cf the continuous spontaneous localisation model of quantum measurement; Pearle, 1989). This linkage is tantalisingly close, but the missing $\sqrt{-1}$ in the transformation is fundamental; the complex form of the Schr\"{o}dinger equation just cannot be obtained by a Cole-Hopf transformation from an underlying real-number Euler-like equation. 

However, this does not invalidate the claim that complex Schr\"{o}dinger dynamics obtains from some underlying real-number non-computable Euler-like dynamics. A certain differentiability is implied in order to apply the Cole-Hopf  transformation, and hence is only really an appropriate description of the transformation between a (smooth) Galerkin-$n$ truncation of the Euler equation. There is an important point and subtle point to be made here. Non-computability properties can be obscure at the level of the differential equation.The `$0$' on the right hand side of the Euler equation may not be identically zero but may represent some (irreversible) process which has only weakly converged to zero in the limit where the Euler equation is relevant.  Analytic transformations such as Cole-Hopf can be blind to this subtlety and (based on the analysis in this paper) can therefore lead to transformed equation sets which are right in form, but where terms are wrong up to a factor of $i$. 

Ultimately this may suggest a limitation in the use of differential equations to describe the fundamental laws of physics. 

\section{Ontology}
\label{sec:ontology} 

In this paper, we have proposed an objective real-number ($r_p$) definition of the quantum state  and a deterministic (`no dice') model of state measurement: $\mathbf{1}$ if $r_p \ge 1/2$. This definition coexists with a dual isomorphic representation of the quantum state in terms of sequences of solutions to recursive approximations to the measurement problem - the latter having a natural complex structure. In this section we discuss some of the fundamental conceptual problems of quantum theory in the light of this duality of representation. 

Consider for example, the 2-slit experiment . If we measure the state $r_p$ at $p \in S$ using a detector close to one slit we will get a definite answer: $\mathbf{1}$, implying a particle detected, or $\mathbf{-1}$, implying a particle not detected. Of course, a  null measurement (ie giving $\mathbf{-1}$) in no way implies that $r_p$ itself is zero (the real $0 \le r\le 1$ is zero with measure zero!).  A measurement by a detector at the other slit is equivalent to a measurement on the state at the antipodal point on $S$; if the first measurement is $\mathbf{1}$, the other will be $\mathbf{-1}$, and vice versa.

We could now ask: suppose we had performed measurements using one of these detectors, on a set of particles that in reality had produced an interference pattern on some background screen, then would the outcome of measurement definitely have been $\mathbf{1}$ (detected) or $\mathbf{-1}$ (not detected)? This is an example of a counterfactual proposition associated with a rotation of the coordinates on our version of the Riemann sphere, mapping computable states to non-computable states. In other words if the physically-realisable states $r_p$ are  associated with the computable reals, then, in general, the states $r^{\prime}_p$ associated with counterfactual measurements  correspond to non-computable reals. Hence, whilst a Turing machine can solve $\mathcal{P}:r_p\ge 1/2$, there is no Turing machine which can determine the truth value of  the proposition $\mathcal{P^{\prime}}: r^{\prime}_p \ge 1/2$.

Now, whilst there is certainly no dispute about  the absolute nature of the truth, or otherwise, of a $\Pi_1$ sentence which asserts that  a certain Turing machine terminates (Penrose, 1994),  it is legitimate to ask whether $\mathcal{P}^{\prime}$ for non-computable $r^{\prime}_p$ is unambiguously either true or false. In essence, as discussed in section \ref{sec:argand}, we are asking under what conditions it makes sense to say that the non-computable real $r^{\prime}_p$ does exist, given that none of its digits in a binary expansion, not even its first one,  is definable by the application of the rules of some first-order formal system. In other words, whilst $\Pi_1$ propositions may or may not be recursively solvable, the real $r^{\prime}_p$ is not even recursively enumerable. 

Let us take the view that $\mathcal{P^{\prime}}$ for non-computable $r_{p^{\prime}}$ do not have definite truth values (a  refinement of the view expressed in Palmer, 1995). Such an approach implies that all physically-realisable measurements are necessarily  associated with states $r_p$ that are computable (else we would have the contradiction that  mathematical propositions associated with physically-realisable experiments do not have definite truth values). This non-Platonist approach is not inconsistent with G\"{o}del's theorem. (G\"{o}del's theorem asserts that the  lack of provability of some  proposition $\mathcal{P}$ about recursively-enumerable sentences, does not imply that $\mathcal{P}$ has no definite truth value. However, as mentioned, we are dealing here with propositions $\mathcal{P}^{\prime}$ about numbers which are not even recursively enumerable.  (In this respect, it is, perhaps, ironic to note that the classic debates about physical reality in quantum theory, between Einstein, Bohr and others,  occurred contemporaneously with, but independently of, the seminal debates on mathematical existence between Brouwer, Hilbert, G\"{o}del and others.)

The author believes that proposing such a linkage between  mathematical and physical existence, whilst certainly unusual,  has both consistency and appeal in a deterministic cosmos where counterfactual experiments are inconsistent with either the laws of physics or the cosmological initial state. It therefore seems appropriate that the outcomes of such experiments should be utterly inaccessible by mathematical analysis. The implications of adopting this approach are immediate and dramatic. For example, it cannot be inferred that quantum states which create interference patterns in a two-slit  experiment would have been associated with localised particles if suitable measurements were to have been made near one or other of the individual slits. In the proposed theory, the states that in reality produce diffraction patterns are by definition computable states, and the corresponding states associated with (counterfactual) measurement at a slit are non-computable. Similarly, the states that in reality are detected close to one or other of the slits are the computable states, and the corresponding states associated with (counterfactual) measurement at the diffraction screen are non-computable.  As such there is no  requirement in this theory for `many worlds'. The theory of course generates a (very!) large number of approximate recursively-solvable worlds, and these can be used to assign computable probabilities to any specified counterfactual, but these recursive worlds are individually fictitious; none corresponds to the real world!

Moreover, without counterfactual definiteness, the model is not constrained to satisfy Bell's inequalities (Palmer, 1995). As such, the proposed model is certainly not classical (a point that has been  stressed in the analysis of  the regularity of weak solutions to the Euler equation), and does not fall into the class of conventional (ie computable) local hidden- variable theories (as an Euler- equation  model using any one of the Galerkin-$n$ truncations would certainly have to be). Hence whilst it is certainly the case that the system proposed here should be distinguised from a classical or conventional local hidden-variable model (it has to be, to be consistent with the quantum correlation $C(\Delta\theta)=-\cos\Delta\theta$), there is no obvious sense in which the proposed model is subject to `spooky action at a distance'.

This `realistic'  interpretation of the quantum state does, at some fundamental level,  appear to restrict the possible measurements that can be performed on a quantum state; we are not free to choose measurement orientations for which the quantum state would be non- computable. In some sense, the author is  arguing that whatever happens in the cosmos is subject to a global constraint that actual physical processes are computable. Like global energy-momentum conservation, this constraint is not localisable, nevertheless, it is not one that endangers the concept of causality. In a deterministic cosmos, the choice of remote EPR measurement orientations, and the entangled quantum state itself, are both deterministically related by common cosmic initial conditions, where relative computational consistency is presumably ensured. 

Whilst this may appear to conflict with the notion of free will, in practice the theory puts no practical constraints on our ability to excercise choice. As such, the strong instinctive attraction we have towards the notion of free will could derive more from Darwinism than from fundamental physics (Hawking, 1993). (Failing to check the road for traffic before crossing it - on the basis that everything is predetermined - would not be conducive to the propagation of the species!). Indeed, the sense of discomfort one may has about determinism (eg from the perspective of moral philosophy) might indicate, rather than some undetected logical flaw in the concept, an evolutionary preconditioning against acceptance of the concept!

However, if there is no inconsistency between underlying determinism and a practical belief in free will, then it should be possible to represent the state vector in a consistent form where there is no mathematical distinction between the actual and the counterfactual (should we choose to do so!). In this paper, we have discussed the dual representation of a non- computable real as some divergent sequence of computable solutions to the measurement criterion,  (compactly) expressed as a complex linear superposition of observable outcomes (either equation \ref{eq:add1} or equation \ref{eq:dirac}). However, the price to pay for adopting this picture of the quantum state  is apparent non-locality, many possible worlds, and a very fuzzy picture indeed of objective reality. The author is clear about which view he prefers! 

\section*{References}

\begin{description}

\item Arnold, V.I. and Khesin, B.A., 1998: Topological methods in hydrodynamics. Applied Mathematical Sciences, 125. Springer-Verlag. New York. 374pp.

\item Duane, G., 2001: Violations of Bell's inequality in synchronised hyperchaos. Foundations of Physics Letters. Accepted. 

\item Frisch, U., 1995: Turbulence: the legacy of A.N.Komolgorov. Cambridge University Press. Cambridge, 296pp

\item Hardy, G.H. and Wright, E.M., 1938: The Theory of Numbers. Oxford University Press. 

\item Hawking, S.W., 1993: Black Holes and Baby Universes. Bantam Press. 

\item Lorenz, E.N., 1969: The predictability of a flow which possesses many scales of motion. Tellus, 21, 289-307.

\item Kraichnan, R.H., 1967: Inertial ranges in two-dimensional turbulence. Phys. Fluids, 10, 1417-1423.

\item Nicolis, J.S., Nicolis, G. and Nicolis, C., 2001: Nonlinear dynamics and the two-slit delayed experiment. Chaos, Solitons and Fractals, in press. 

\item Ott, E., Sommerer, J.C., Alexander, J.C., Kan, I., and Yorke, J.A., 1994: Scaling behavior of chaotic systems with riddled basins. Phys.Rev.Lett., 71, 4143-4137.

\item Palmer, T.N., 1995: A local deterministic model of quantum spin measurement. Proc.R.Soc.Lond. A, 451, 585-608.

\item Pearle, P., 1989: Combining stochastic dynamical state vector reduction with spontaneous localisation. Phys.Rev., A39, 2277-2289.

\item Penrose, R., 1994: Shadows of the mind. Oxford University Press. Oxford. 457pp

\item Scheffer, V., 1993: An inviscid flow with compact support in spacetime. J.Geom.Analysis, 3, No 4, 343-401.

\item Shnirelman, A., 1997: On the nonuniqueness of weak solutions to the Euler equation. Comm. Pure \& Appl. Math., 50, 1260-1286.

\item Shnirelman, A., 2001: Weak solutions with decreasing energy of the incompressible Euler equations. Comm.Pure \& Appl. Math., in press.

\item Sommerer, J.C. and Ott, E., 1993: A physical system with qualitatively uncertain dynamics. Nature, 365, 138-140.

\item Whitham, G.B., 1974: Linear and Non-linear waves. John Wiley. New York. 636pp.

\end{description}
\end{document}